\documentclass[12pt]{article}

\usepackage{graphicx}
\usepackage{amsfonts}
\newtheorem{proposition}{Proposition}

\newtheorem{definition}[proposition]{Definition}

\def\+{{+\!\!\!+}}

\def\d{\partial}

\def\pmb#1{\setbox0=\hbox{#1}% 
\kern.0em\copy0\kern-\wd0 
\kern-.04em\copy0\kern-\wd0 
\kern.08em\copy0\kern-\wd0 
\kern-.04em\raise.0433em\box0 }         %poor man's bold macro (TexBook) 
\newcommand{\nc}{\newcommand} 
\nc{\beq}{\begin{equation}} 
\nc{\eeq}[1]{\label{#1}\end{equation}} 
\nc{\ber}{\begin{eqnarray}} 
\nc{\eer}[1]{\label{#1}\end{eqnarray}} 
\nc{\pek}[1]{\cite{#1}} 
\nc{\enr}[1]{(\ref{#1})} 
\nc{\kal}[1]{{\cal{#1}}} 
\nc{\dott}{\;\cdot\;} 
\newcommand{\Section}[1]{\section{#1} \setcounter{equation}{0}}

\def\0 {\nonumber}

\begin{document} 
\setcounter{page}{0}
\newcommand{\inv}[1]{{#1}^{-1}} %inverse 
\renewcommand{\theequation}{\thesection.\arabic{equation}} 
\newcommand{\be}{\begin{equation}} 
\newcommand{\ee}{\end{equation}} 
\newcommand{\bea}{\begin{eqnarray}} 
\newcommand{\eea}{\end{eqnarray}} 
\newcommand{\re}[1]{(\ref{#1})} 
\newcommand{\qv}{\quad ,} 
\newcommand{\qp}{\quad .} 
\thispagestyle{empty}
%\begin{titlepage} 
%\title{} 

\begin{flushright} \small
 SISSA 47/2008/FM \\  
 UUITP-16/08\\
 NSF-KITP-08-117\\                        
\end{flushright}
\smallskip
\begin{center}\LARGE
                         
 { \bf Topological branes, p-algebras and \\ generalized Nahm equations}
  \\[12mm] \normalsize
{\bf Giulio Bonelli$^{a}$}, {\bf Alessandro Tanzini$^a$} 
and {\bf Maxim Zabzine$^{b}$,}\\[8mm]
 {\small\it
$^a$International School of Advanced Studies (SISSA) and INFN, Sezione di Trieste\\
 via Beirut 2-4, 34014 Trieste, Italy
~\\
~\\
$^b$Theoretical Physics, Department of Physics and Astronomy, \\
Uppsala University,
Box 803, SE-751 08 Uppsala, Sweden }
\end{center}
\vspace{10mm}
\centerline{\bf\large Abstract}
\bigskip
\noindent
 Inspired by the recent advances in multiple M2-brane theory, we consider the generalizations of
 Nahm equations for arbitrary $p$-algebras. We construct the topological $p$-algebra 
 quantum mechanics associated to them and we show that this can be obtained as a truncation of the 
 topological $p$-brane theory previously studied by the authors. 
 The resulting topological $p$-algebra quantum mechanics is discussed in detail and 
 the relation with the M2-M5 system is pointed out in the $p=3$ case, providing a geometrical argument 
 for the emergence of the $3$-algebra structure in the Bagger-Lambert-Gustavsson theory.
\eject
\normalsize

\eject
 
\section{Introduction}

Recently the appearance of new classes of gauge theories stimulated 
a growing interest in $p$-algebras as possible generalizations of the standard Lie algebras
in the description of gauge interactions. 
In particular, based on a $3$-algebra, a multiple M2-brane theory \cite{Bagger:2006sk, Gustavsson:2007vu} 
has been constructed overcoming the no-go theorem \cite{Schwarz:2004yj} 
on the formulation of superconformal ${\cal N}=8$ Chern-Simons theories.
A crucial input for this construction came from the study of the M2-M5 system
in the Basu-Harvey's work \cite{Basu:2004ed} where an equation describing the BPS bound state of multiple 
M2-branes ending on an M5 was formulated.
The Basu-Harvey equation is a generalization of the Nahm equation
\cite{Nahm:1979yw} involving an algebraic structure modeled on the Nambu bracket
\cite{Nambu:1973qe} (for a recent review see \cite{Curtright:2002fd}). $p$-algebras 
can be obtained as linearizations of Nambu algebras 
\cite{Takhtajan:1993vr}. This has been used in \cite{chongsong} to infer the 
$N^{3/2}$ scaling of the degrees of freedom of the M2-branes (see also \cite{anche} for a different argument).
A wide amount of literature has been inspired on related subjects \cite{tutti}.

In a series of
  papers \cite{Bonelli:2005ti, Bonelli:2005rw, Bonelli:2006ph}, the authors formulated a 
class of topological theories for $p$-branes. 
 These are based on the realization of $p$-brane instantons wrapping calibrated cycles
as solution of BPS bounds for the Nambu-Goto action.  We will observe that
these equations can be rewritten in terms of Nambu brackets. In this note
  we present a $p$-algebra version of these topological theories
based on the correspondence between the Nambu brackets and $p$-algebras. 

In particular, for $p=3$ we will show that the discretization of the 3-brane instanton coincides with the generalized 
Basu-Harvey equation as studied in \cite{Berman:2006eu, Krishnan:2008zm}.
This corresponds to M2-M5 system compactified on 
%$R^{1,2}\times X$, with $X$ 
a $Spin(7)$ holonomy manifold, with
%the M2's along $R^{1,2}$ and 
the M5 wrapping a 
%along $R^{1,1}\times C$ with $C$ a 
calibrated Cayley four-cycle.
Therefore, in the internal manifold, the M5 
%results in a $3$-brane wrapping the four cycle $C$
%and as such 
can be described in terms of a topological $3$-brane theory.
From the M2-brane viewpoint, this can be recovered as the large $N$ limit of the generalized Basu-Harvey equation.
This provides a geometrical argument for the emergence of the $3$-algebra structure in the Bagger-Lambert theory.
Moreover, our classification of topological $p$-brane theories and the corresponding calibrations encompasses all 
the intersecting M2-M5 brane systems studied in \cite{baggerwest}.

The paper is organized as follows. 
In Section \ref{topbrane} we review the construction of topological 
 $p$-brane theory associated to a real (complex) cross vector product.  
Section \ref{matrix} is devoted to the formulation of the $p$-algebra version of the topological $p$-brane theories.
In Section \ref{examples} we provide some examples and applications of our formalism. We discuss 
 the $p=2$ case corresponding to the description of membrane instanton via generalized Nahm equations 
and the $p=3$ case corresponding to the relation between 3-algebras and M2-M5 systems.
Section \ref{end} is left for concluding comments and open questions.
 In order to make this paper more readable we collect in 
Appendix \ref{VCP} some facts about  the cross vector products   
and in 
Appendix \ref{Nambu} we  review the Nambu brackets and (Nambu-Lie) $p$-algebras.

\section{Topological brane theory}
\label{topbrane}

In this section we review the topological p-brane theory elaborated in \cite{Bonelli:2005ti, Bonelli:2005rw, Bonelli:2006ph}.  
This is a cohomological field theory which
generalizes
the Baulieu-Singer approach \cite{Baulieu:1989rs} to topological two dimensional sigma model (A-model)
to topological branes associated with real (complex) vector cross products.
Indeed the present construction can be naturally interpreted within the Mathai-Quillen formalism.
    
Consider the Nambu-Goto $p$-brane theory on the manifold $M$ with Riemannian metric $g$ 
 \beq
 S = \int d^{p+1}\sigma\,\,\sqrt{\det(\d_\alpha X^\mu g_{\mu\nu} \d_\beta X^\nu)}~,
 \eeq{Naameppp}
  where $\alpha=0,1,...,p$ and $X$ are the maps from $(p+1)$-dimensional world-volume 
   $\Sigma_{p+1}$ to $M$. 
   If $\Sigma_{p+1} = I \times \Sigma_p$ with $I$ being the interval ($S^1$ or $\mathbb{R}$) we can 
    gauge fix diffeomorphisms and obtain the following action
      \beq
    S = \frac{1}{2} \int d^{p+1}\sigma\,\,\left ( \dot{X}^\mu g_{\mu\nu} \dot{X}^\nu +
       \det (\d_a X^\mu g_{\mu\nu} \d_b X^\nu) \right )~,
\eeq{gehak12903}
  with $\dot{X}^\mu =\d_0 X^\mu$ and $a,b=1,...,p$ label the directions along $\Sigma_p$.  
   Assuming that there is a $(p+1)$-form $\phi$ on $M$ we can write down the bound
  \beq
\int d^{p+1}\sigma\,\, \left ( \dot{X}^\mu \pm \phi^\mu_{\,\,\,\nu_1...\nu_p} \d_1 X^{\nu_1}...
 \d_p X^{\nu_p} \right ) g_{\mu\lambda}
 \left ( \dot{X}^\lambda \pm \phi^\lambda_{\,\,\,\sigma_1...\sigma_p}\d_1 X^{\sigma_1}...
  \d_p X^{\sigma_p}
  \right ) \geq 0 ~.
\eeq{boudnao20AAA}      
  If $\phi$ and $g$ correspond to a vector cross product structure on $M$ (see Appendix A for 
   the definition and properties), then the bound (\ref{boudnao20AAA}) can be rewritten as follows
\beq
 \frac{1}{2} \int d^{p+1}\sigma\,\,\left ( \dot{X}^\mu g_{\mu\nu} \dot{X}^\nu +
       \det (\d_a X^\mu g_{\mu\nu} \d_b X^\nu) \right ) \geq \mp \frac{1}{(p+1)!} \int X^*(\phi)~.
\eeq{membrabound1AAA}
 Moreover if $d\phi=0$ the right-hand side is a topological invariant. The bound (\ref{membrabound1AAA})
  is saturated  for
 \beq
 \dot{X}^\mu \pm \frac{1}{p!} \epsilon^{a_1 ... a_p}
 \phi^\mu_{\,\,\,\nu_1 ... \nu_p} \d_{a_1} X^{\nu_1}... \d_{a_p} X^{\nu_p} =0~.
\eeq{generinstanton} 
We will name the solutions of this equation as $p$-brane instantons. 
Actually, one of their properties is that they span submanifolds in $M$
which are calibrated by $\phi$.
 
Let's consider the topological  $p$-brane theory defined by the topological term
 \beq
  S_{top} = - \frac{1}{(p+1)!} \int  X^*(\phi)~,
 \eeq{topactiongen}
  where $\phi$ is a closed $(p+1)$-form corresponding to a cross vector product on $M$.
   We want to construct the gauge fixed action for this theory.  
  In what follows we would consider the case when $\nabla \phi =0$ with $\nabla$ being 
   the Levi-Civita connection, however this is 
   not essential for the construction and everything can be generalized for a generic 
    closed $\phi$ (see \cite{Bonelli:2005rw}).  
 The action (\ref{topactiongen})  is invariant under the gauge symmetry $\delta X^\mu =\epsilon^\mu$. 
  We define the  BRST transformations as follows
    \beq
  sX^\mu = \psi^\mu~,~~~~~~
  s\psi^\mu =0~,~~~~~~
  s\bar{\psi}^\mu = b^\mu~,~~~~~~
  sb^\mu =0~,
  \eeq{BRST1}
   where the fields $X, \psi, \bar{\psi}, b$ have the ghost numbers $0, 1, -1, 0$ correspondingly.
   Choosing the gauge function as
   \beq
    {\cal F}^\mu = \dot{X}^\mu + \frac{1}{p!} \epsilon^{a_1 ... a_p}\phi^\mu_{\,\,\,\nu_1...\nu_p} \d_{a_1} X^{\nu_1}...\d_{a_p} X^{\nu_p}
     + \Gamma^\mu_{\,\,\,\sigma\rho} \bar{\psi}^\sigma \psi^\rho~,
   \eeq{gaugefixfubcgen} 
  we define the gauge fixed action as 
$$ S_{GF} =-\frac{1}{(p+1)!} \int X^*(\phi) + 
 \int d^{p+1}\sigma\,\, s \left ( \bar{\psi}^\mu (g_{\mu\nu} \dot{X}^\nu 
      + \frac{1}{p!} \epsilon^{a_1 ... a_p}\phi_{\mu{\nu_1}...{\nu_p}} \d_{a_1} X^{\nu_1} ...\d_{a_p} X^{\nu_p} +\right . $$
\beq     
\left .       + \frac{1}{2} \Gamma_{\mu\sigma\rho}
       \bar{\psi}^\sigma \psi^\rho - \frac{1}{2} g_{\mu\nu} b^\nu ) \right ) ~.
  \eeq{gaugeficm119}
After path integrating  over $b^\mu$, we arrive to the following gauge fixed action
 $$S_{GF}=  \int d^{p+1}\sigma \left ( \frac{1}{2} \dot{X}^\mu g_{\mu\nu} \dot{X}^\nu
  + \frac{1}{2}\det (\d_a X^\mu g_{\mu\nu} \d_b X^\nu)
     - \bar{\psi}^\mu g_{\mu\nu} \nabla_0 \psi^\nu - \right . $$
\beq
\left . -\frac{1}{(p-1)!}\phi_{\mu{\nu_1}{\nu_2}...{\nu_p}} \bar{\psi}^\mu \nabla_{a_1} \psi^{\nu_1}
      \d_{a_2} X^{\nu_2} ...\d_{a_p} X^{\nu_p}
       \epsilon^{a_1 a_2 ... a_p} + \frac{1}{4} {\cal R}_{\mu\sigma\lambda\rho} \bar{\psi}^\mu
      \psi^\rho \bar{\psi}^\sigma \psi^\lambda \right )~,
  \eeq{gjdkwp293}
  where $\nabla_\alpha \psi$ is defined as
  \beq
  \nabla_\alpha \psi^\mu = \d_\alpha \psi^\mu  + \Gamma^\mu_{\,\,\,\rho\lambda} \d_\alpha X^\rho
   \psi^\lambda~.
  \eeq{nbladef}
The action (\ref{gjdkwp293}) is invariant under the following BRST transformations
\beq
sX^\mu = \psi^\mu~,\,\,\,\,\,\,\,\,\,\,
  s\psi^\mu =0~,\,\,\,\,\,\,\,\,\,\,
  s\bar{\psi}^\mu =  \dot{X}^\mu + \phi^\mu_{\,\,\,\nu_1 ... \nu_p} \d_1 X^{\nu_1}... \d_p X^{\nu_p} + \Gamma^\mu_{\,\,\,\sigma\rho}
   \bar{\psi}^\sigma \psi^\rho ~,
\eeq{BRSTgenrpq}
 which are nilpotent on-shell.   The action (\ref{gjdkwp293})  can be rewritten as
 \beq
  S_{GF} =  - \frac{1}{(p+1)!} \int  X^*(\phi) + \frac{1}{2} \int d^{p+1}\sigma\,\, s\left (
  \bar{\psi}^\mu (g_{\mu\nu} \dot{X}^\nu 
      + \phi_{\mu{\nu_1}...{\nu_p}} \d_1 X^{\nu_1} ...\d_p X^{\nu_p} ) \right ).
 \eeq{fghs03wr888}
 This implies that this cohomological model is localized on $p$-brane instantons
  (\ref{generinstanton}).  The observables are labeled by the de Rham cohomology 
   of $M$ and  the resulting
    TFT produces some invariants associated to a moduli space of $p$-brane instantons
\cite{Bonelli:2006ph}.

  These TFTs are specified by the real vector cross products $(\phi, g)$. Indeed the 
   real vector cross products
   have been classified   by Brown and Gray \cite{brown} (see the list in Appendix). 
    There are four different cases when the cross product exists. The first case corresponds 
   to  $\phi$ being the volume form on $M$. In this case the TFT we constructed corresponds
      to $p$-brane theory embedded into $p+1$ dimensional space $M$.  
       The second case corresponds
        to a symplectic manifold with $\phi$ being a closed non-degenerate $2$-form.
         The  corresponding TFT  is just topological two dimensional 
          sigma model (A-model)
          \cite{Witten:1988xj}.  The remaining two vector cross product structures are the exceptional 
 cases.  The first exceptional case corresponds to seven dimensional manifold $M$ with 
   $G_2$-structure, $\phi$ being the invariant $3$-form. This TFT describes a topological  membrane theory
    localized on the associative maps.
   The second exceptional case corresponds to eight dimensional 
     manifolds with $Spin(7)$-structure where $\phi$ is the associated $4$-form (the Cayley form).
In this case we have a   topological theory for $3$-branes wrapping Cayley submanifolds. 
      
   The above construction can be extended to the case of complex vector products $(\phi, J, g)$. 
As an example, one can consider $\phi$ to be the  holomorphic volume form on a Calabi-Yau manifold
and in this case the topological brane theory localizes on submanifolds calibrated by the real part 
     of the holomorphic volume form. For further details the reader may consult \cite{Bonelli:2005rw}.

 Let us conclude with a few general remarks about these theories.  Except for the A-model, the study of
  these TFTs is obstructed by the lack of knowledge about the moduli space of $p$-brane instantons.
   Moreover, it is difficult to formulate these TFTs in a fully covariant formalism. 
   One possibility for $p$ even is to use the contact structure on $\Sigma_{p+1}$. 
Alternatively one can easily write a covariant version 
   of instanton equation (\ref{generinstanton}), but this provides redundant gauge fixing conditions 
(see \cite{Bonelli:2006ph} for the membrane case).
The discretized version which we will explain  in the next Section may help to overcome 
 some of these difficulties. 
    
\section{$p$-algebra topological brane theories}
\label{matrix}

In this Section we discuss the $p$-algebra version of the topological $p$-brane theories that we just 
reviewed.  
The starting point of our construction is the observation that 
on the spatial part $\Sigma_p$ of the world-volume a canonical  Nambu bracket\footnote{See Appendix \ref{Nambu} for review of the Nambu 
and $p$-algebra brackets.} is defined
through the Jacobian 
\beq
\{ X^{\nu_1}, X^{\nu_2}, ..., X^{\nu_p}\}  
=  \epsilon^{a_1 ... a_p}
  \d_{a_1} X^{\nu_1}... \d_{a_p} X^{\nu_p}.
\eeq{definitNabucoord}
This Nambu bracket appears through the whole construction. In particular 
the topological term (\ref{topactiongen})  can be rewritten as
\beq
 S_{top} = -\frac{1}{p!}~ \int d^{p+1}\sigma ~ \dot{X}^{\nu_0} \{ X^{\nu_1}, X^{\nu_2}, ..., X^{\nu_p}\}  ~\phi_{\nu_0 \nu_1 ... \nu_p}
\eeq{newtopteNambu}
 and the $p$-brane instanton (\ref{generinstanton})
\beq
  \dot{X}^\mu + \frac{1}{p!}~\phi^\mu_{~~\nu_1\nu_2 ...\nu_p} \{ X^{\nu_1}, X^{\nu_2}, ..., X^{\nu_p}\}=0~.
\eeq{gaugefixingNambu}
 The idea is to replace everywhere the  Nambu bracket by the corresponding $p$-algebra 
  bracket in a consistent way. 
    In more physical terms this formal replacement can be viewed as some sort of 
   truncation of the $p$-brane theory, very much in the spirit of the original matrix membrane 
    theory \cite{de Wit:1988ig}. 
    From the mathematical point of view $p$-algebra brackets can be regarded as a linearization of Nambu brackets  \cite{Takhtajan:1993vr}. 

Let us perform the systematic analysis of the model. 
Consider a theory on the interval (circle) $I$ with the bosonic field  $X(t)$  taking value in  $\mathbb{R}^d \otimes V$, 
where $V$ is some representation of the $p$-algebra. 
We use the following convention $X^\mu(t) = X^\mu_a  (t)T^a$ with $\mu$ being an index along $\mathbb{R}^d$ and $T^a$'s form a basis in $V$. 

Assume now that  $\mathbb{R}^d$ 
 is equipped with a constant vector cross product\footnote{In the whole discussion $\mathbb{R}^d$ can be replaced by any space where  
we can define a constant cross vector product, e.g. torus $\mathbb{T}^d$ etc.}  $(\phi, g)$.  Then in this setup 
the $p$-brane instanton equation (\ref{gaugefixingNambu}) becomes
  \beq
  \dot{X}^\mu + \frac{1}{p!}~\phi^\mu_{~~\nu_1\nu_2 ...\nu_p} [ X^{\nu_1}, X^{\nu_2}, ..., X^{\nu_p} ]=0~,
\eeq{gaugefixingalgebra}
 which is a generalization of the original Nahm equation \cite{Nahm:1979yw} to the $p$-algebra case
 (for earlier discussion of such generalization for $p=3$ see the  Basu-Harvey's work \cite{Basu:2004ed}). 
 We will review some examples of (\ref{gaugefixingalgebra}) in the next Section.  
If the $p$-algebra $V$ is equipped with an invariant inner product\footnote{As far as our classical discussion is concerned, 
we do not require the inner product to be positive definite. However, notice that if one wants to regard the $p$-algebra as coming 
as a truncation of the Nambu algebra of functions on $\Sigma_p$, then one should keep this requirement.}, which we denote by ``$Tr$'', 
then the topological term (\ref{newtopteNambu}) becomes 
\beq
  S_{top} = - \frac{1}{p!}~\int dt ~ Tr \left (\dot{X}^{\nu_0} [X^{\nu_1}, X^{\nu_2}, ..., X^{\nu_p} ] \right )~\phi_{\nu_0 \nu_1 ... \nu_p} ~.
\eeq{topologicaltermmatrix}
 This term is invariant under $\delta X^\mu = \epsilon^\mu$, modulo appropriate boundary conditions 
  if we consider the theory on the interval. 

 Introducing now the fields $\psi$, $\bar{\psi}$, 
  $b$ valued in $\mathbb{R}^d \otimes V$ with ghost numbers $1, -1, 0$ respectively, we define
the  BRST symmetry $s$ in the same way as in (\ref{BRST1}).
    In analogy with the previous discussion the gauge fixed
 action is
 \beq
  S_{GF} = S_{top} + s \int dt ~Tr \left ( \bar\psi^\mu, (\dot{X}_\mu + \frac{1}{p!} \phi_{\mu{\nu_1}...{\nu_p}} [X^{\nu_1}, ... , X^{\nu_p} ] - \frac{1}{2} b_\mu ) \right )~,
 \eeq{matrixgafixing}
  where we use a constant metric $g$ on $\mathbb{R}^d$ to contract the indices.
 Using the BRST transformations and integrating out $b$ we get
 $$ S_{GF} = \int dt~ Tr \left ( \frac{1}{2} \d_0 X^\mu \d_0 X_\mu - \bar{\psi}^\mu \d_0 \psi_\mu 
   + \frac{1}{2p!} [X^{\nu_1}, ..., X^{\nu_p}] [X_{\nu_1}, ... ,X_{\nu_p} ]  -  \right .$$
   \beq
  \left . - \frac{1}{(p-1)!} 
   \bar{\psi}^{\mu_0} [\psi^{\mu_1}, X^{\mu_2}, ... , X^{\mu_p}] \phi_{\mu_0 \mu_1 ... \mu_p} \right )~.
 \eeq{gaugefixactionmatrixst}
  %Expanding in the basis $T^a$ in $V$ we can rewrite above action as follows
  %$$ S_{GF} = \int dt~  \left ( \frac{1}{2} \d_0 X^{a\mu} \d_0 X_{a\mu} - \bar{\psi}^{a\mu} \d_0 \psi_{a\mu} 
  % + \frac{1}{2p!} Tr ( [X^{\nu_1}, ..., X^{\nu_p}] [X_{\nu_1}, ... ,X_{\nu_p} ] ) -  \right .$$
  % \beq
  % \left . - \frac{1}{(p-1)!} 
  % \bar{\psi}_{a_1}^{\mu_0} \psi^{\mu_1}_{a_2} X^{\mu_2}_{a_3} ... , X^{\mu_p}_{a_p} f^{a_1 a_2 ... a_p} \phi_{\mu_0 \mu_1 ... \mu_p} \right )~.
 % \eeq{gaugefixactionmatrixstindex}
  This action is reminiscent of the Bagger-Lambert action for 3-algebras \cite{Bagger:2006sk}.
 In the next Section we will provide more details on the BPS equation of the Bagger-Lambert-Gustavsson theory which corresponds to the 3-algebra version of the
topological $3$-brane instantons.
 
   The action (\ref{gaugefixactionmatrixst})  has the following odd symmetries
   \beq
    sX^\mu = \psi^\mu~,~~~~~~s\psi^\mu = 0~,~~~~~~s\bar{\psi}^\mu =  \d_0 {X}^\mu + \frac{1}{p!}\phi^\mu_{~~\nu_1\nu_2 ...\nu_p} 
[ X^{\nu_1}, X^{\nu_2}, ..., X^{\nu_p} ]
   \eeq{matrixstran}
  \beq
    \bar{s}X^\mu = \bar{\psi}^\mu~,~~~~~~\bar{s}\psi^\mu = \d_0 {X}^\mu - \frac{1}{p!} \phi^\mu_{~~\nu_1\nu_2 ...\nu_p} 
[ X^{\nu_1}, X^{\nu_2}, ..., X^{\nu_p} ] ~,~~~~~~\bar{s}\bar{\psi}^\mu =  0  
 \eeq{matrixstranbar}
 which obey the following on-shell algebra
 \beq
  s^2=0~,~~~~~~~\bar{s}^2=0~,~~~~~~~ s\bar{s} + {\bar s} s = 2 \d_0~.
 \eeq{algebraN2matrix}
  Thus the action (\ref{gaugefixactionmatrixst}) displays $N=2$  supersymmetry.  
Moreover, (\ref{gaugefixactionmatrixst}) is invariant under the adjoint action of the $p$-algebra
\beq
\delta_\Lambda X^\mu=ad_\Lambda X^\mu,~~~~~~ \delta_\Lambda \psi^\mu= ad_\Lambda \psi^\mu,~~~~~~~
\delta_\Lambda \bar\psi^\mu= ad_\Lambda \bar\psi^\mu
\eeq{adjointact}
where $\Lambda \in \wedge^{p-1} V$ is a constant parameter
and $ad_\Lambda v^\mu_a=f^{b_1.. b_{p}}_{~~~~~~a}  \Lambda_{b_1 ... b_{p-1}}v_{b_p}^\mu$
(see Appendix B for further details).
This symmetry is remnant of the spacial volume-preserving infinitesimal 
 diffeomorphisms  on $\Sigma_p$ 
 after the truncation to the $p$-algebra.
Indeed the adjoint action is a derivation  of $p$-algebra and gives rise to  a Lie algebra $\left[\delta_\Lambda,\delta_{\Lambda'}\right]=\delta_{[[\Lambda,\Lambda']]}$. 
 This is similar to the case of Nambu bracket  where 
  the infinitesimal volume-preserving diffeomorphisms
  are derivation of the Nambu bracket and they form a Lie algebra.
%  \beq
%   \delta_\Lambda X^\mu_a = f^{b_1.. b_{p}}_{~~~~~~a}  \Lambda_{b_1 ... b_{p-1}} X_{b_p}^\mu~,~~~~~~
%    \delta_\Lambda \psi^\mu_a = f^{b_1.. b_{p}}_{~~~~~~a}  \Lambda_{b_1 ... b_{p-1}} \psi_{b_p}^\mu~,~~~~~~
%     \delta_\Lambda \bar{\psi}^\mu_a = f^{b_1.. b_{p}}_{~~~~~~a}  \Lambda_{b_1 ... b_{p-1}} \bar{\psi}_{b_p}^\mu~,
%  \eeq{adjointact}
 Since the adjoint action is a derivation of the bracket we have 
 \beq
  s \delta_\Lambda - \delta_\Lambda s =0~,~~~~~~~~~~~~~~~~
  \bar{s} \delta_\Lambda - \delta_\Lambda \bar{s} =0~.
 \eeq{commutingsymm}
%  Therefore there are additional nilpotent symmetries of the action 
%  \beq
%  s_\Lambda = s \delta_\Lambda~,~~~~~~~~~~~~~~~~
%  \bar{s}_\Lambda = \bar{s} \delta_\Lambda~.
% \eeq{additionalnilpot}
%  It is important to stress that the topological $p$-brane theory considered in the previous Section 
%   does not have these additional symmetries.
In fact the action (\ref{gaugefixactionmatrixst}) can be rewritten in superfield formalism 
 where $N=2$ supersymmetry becomes manifest. Namely introducing the superfield 
$\mathbb{X}: I^{1|2} \rightarrow \mathbb{R}^d \otimes V$ written in components as
$$\mathbb{X} = X + \theta^+ \frac{1}{\sqrt{2}}(\psi + \bar{\psi}) + \theta^- 
\frac{1}{i\sqrt{2}}(\psi - \bar{\psi}) + \theta^+ \theta^- F$$
and the covariant odd derivatives 
$$ \mathbb{D}_\pm = \frac{\d}{\d \theta^\pm} + \theta^\pm \d_0~,$$
we can write the action
\beq
 S_{GF} = - \frac{1}{2} Tr \int dt ~d\theta^+ d \theta^- ~  \left (\mathbb{D}_+ \mathbb{X}^\mu
  \mathbb{D}_- \mathbb{X}_\mu + W(\mathbb{X})
 \right )
\eeq{susyactionfield}
 with the superpotential
\beq
 W(\mathbb{X}) = - \frac{2i}{(p+1)!} f^{a_0 a_1 ... a_p} \phi_{\nu_0 \nu_1 ... \nu_p} \mathbb{X}_{a_0}^{\nu_0}
 \mathbb{X}_{a_1}^{\nu_1} ... \mathbb{X}_{a_p}^{\nu_p}~.
\eeq{superpotential}
 Upon integrating over the odd measure and the auxiliary field $F$ the action (\ref{susyactionfield})
  produces the component version (\ref{gaugefixactionmatrixst}).  
   From the point of view of additional supersymmetries the action (\ref{susyactionfield}) is somehow
     very amazing. Namely, any transformation of the form 
     \beq
      \delta \mathbb{X} = \mathbb{J} (\epsilon^+ \mathbb{D}_+ + \epsilon^- \mathbb{D}_- ) \mathbb{X}~,
     \eeq{additionalsusy}
      with $\mathbb{J} : \mathbb{R}^d \otimes V \rightarrow  \mathbb{R}^d \otimes V $ gives 
       rise to a (pseudo)-supersymmetry\footnote{The supersymmetry transformations square to translations and pseudo-supersymmetry 
to minus translations. Since we are not concerned with the hermiticity properties here we may consider both cases.} algebra if $\mathbb{J}^2 = \pm 1$. 
The kinetic 
        term in the action is invariant if the metric and $Tr$ are hermitian with respect to $\mathbb{J}$. 
         Indeed this free kinetic term is the most supersymmetric action one can construct. 
         The superpotential (\ref{susyactionfield}) $W$ should be invariant by itself.  In general it is 
          hard to construct superpotentials invariant under a large amount of supersymmetry 
           and the ability to write such superpotentials is related to the existence of invariant tensors
            with respect to rotations. For example, if we look at the case 
             $\mathbb{R}^8 \otimes \mathbb{R}^4$  with $\mathbb{R}^4$ equipped with the canonical $3$-bracket (given by epsilon-tensor) 
and $\mathbb{R}^8$ with Cayley $4$-form $\Psi$ then the superpotential 
             $$ W(\mathbb{X}) = - \frac{i}{12} \epsilon^{a_0 a_1 a_2 a_3} \Psi_{\nu_0 \nu_1 \nu_2 \nu_3} \mathbb{X}_{a_0}^{\nu_0}
 \mathbb{X}_{a_1}^{\nu_1}  \mathbb{X}_{a_2}^{\nu_2} \mathbb{X}_{a_3}^{\nu_3}$$
  has $SO(4)\times Spin(7)$ symmetry group. 
  Notice that this generalizes the superpotential appearing in the ${\cal N}=2$ formulation of the 
  Bagger-Lambert-Gustavsson theory given in \cite{klebanov}. 
   It is worthwhile to remark that the quantum mechanical model (\ref{susyactionfield}) 
   admits easily higher dimensional generalizations. 
      
    Coming back to (\ref{gaugefixactionmatrixst}),
 the global symmetry (\ref{adjointact}) can be gauged by
introducing a connection $A\in\wedge^{p-1}V$
%= A_{a_1 ... a_p} T^{a_1} \wedge ... \wedge T^{a_p}$  
which transforms as
  \beq
   \delta_\Lambda A = - \d_0 \Lambda - [[A, \Lambda]]~,
  \eeq{transformagaugefiel}
   where the bracket $[[~,~]]$ is understood as a bracket of derivations. 
   The corresponding gauge invariant action is
    $$ S_{GF} = \int dt~ Tr \left ( \frac{1}{2} D_0 X^\mu D_0 X_\mu - \bar{\psi}^\mu D_0 \psi_\mu 
   + \frac{1}{2p!} [X^{\nu_1}, ..., X^{\nu_p}] [X_{\nu_1}, ... ,X_{\nu_p} ]  -  \right .$$
   \beq
  \left . - \frac{1}{(p-1)!} 
   \bar{\psi}^{\mu_0} [\psi^{\mu_1}, X^{\mu_2}, ... , X^{\mu_p}] \phi_{\mu_0 \mu_1 ... \mu_p} \right )~,
 \eeq{gaugefixactionmatrixstingdhs}
 which is invariant under the transformations (\ref{adjointact}) and (\ref{transformagaugefiel})
  with now $\Lambda$ being an arbitrary function.
  Correspondingly the covariant derivatives are  defined as
  \beq
   D_0 X^\mu= \d_0 X^\mu + ad_A X^\mu~,~~~~~~~~~~~~~~~~
   D_0 \psi^\mu = \d_0 \psi^\mu + ad_A \psi^\mu~.
  \eeq{defincovarder}
   Alternatively we could use the matrix conventions for the adjoint action, see Appendix \ref{Nambu}. 
 The action (\ref{gaugefixactionmatrixstingdhs}) in the
     case of $3$-algebra could be obtained through an appropriate truncation of the Bagger-Lambert
 action \cite{Bagger:2006sk}. 
   
   \section{Examples and applications}
 \label{examples}
 
 In this Section we briefly go through some examples and applications of the p-algebra theories we have 
 constructed. 

We start with the $p=2$ case where we discuss the relation between membrane instantons, Nahm equations and their generalization 
to octonion algebra.
We then move to the topological 3-brane case: as we discussed in the introduction this topological theory is 
effectively counting
solutions to the Basu-Harvey equations.

%We finally discuss some general features of p-algebra theories.
 %  The main message we would like to convey is that there is an intricate relation 
  %   between different TFTs. Moreover the matrix version of topological $p$-brane theories
   %   may help to calculate many things in the full $p$-brane theory. 
 
 \subsection{Topological membranes and Nahm equation}
 
 Consider the membrane theory on $\mathbb{R}^3$
  with a corresponding cross vector product given by the volume form $\epsilon_{\mu\nu\rho}$. 
 This theory describes maps $X:S^1\times\Sigma_2\to \mathbb{R}^3$ and is localised on those satisfying
 \beq
    \d_0 X^\mu +  \epsilon^\mu_{~\nu\rho}~  \d_1 X^\nu  \d_2 X^\rho = 0~.
 \eeq{local3dimbr}
  It has been pointed out already in 1989 \cite{Ward:1989nz} 
   that equation (\ref{local3dimbr}) is related to the Nahm equation (also see \cite{Curtright:1997st}
    for  related interesting discussions). 
 The topological term of the corresponding matrix theory is 
  \beq
   S_{top} = - \frac{1}{2} \int dt~ Tr \left ( \dot{X}^\mu [ X^\nu, X^\rho ] \right ) \epsilon_{\mu\nu\rho}~.
  \eeq{quantopterm}
 The gauge fixed theory is localized on the solutions of 
   the well-known Nahm equation
 \beq
  \frac{dX^\mu}{dt} + \frac{1}{2} \epsilon^\mu_{~\nu\rho}~ [X^\nu, X^\rho] = 0~. 
 \eeq{standardNahm}
Indeed the large $N$-limit of this topological matrix theory should reproduce the calculations 
of the membrane theory above.
It is worthwhile to remark that (\ref{standardNahm}) is the 
dimensional reduction to one dimension of the self-duality equation for a gauge field in four dimensions.
The invariant (\ref{quantopterm}) comes from $\int_{M_4} Tr (F \wedge F)$.
Thus this theory can be regarded as a reduction to one dimension of 
Donaldson-Witten theory \cite{Witten:1988ze, Baulieu:1988xs}. 
 
% \subsection{Associative Nahm equation}
 
The above construction can be extended to higher dimensional target spaces, 
namely to seven dimensional $G_2$ manifolds with a constant cross vector product.
In this case the theory localizes on associative maps, whose   
% The topological membrane theory on seven dimensional $G_2$ manifold was the main 
%  motivation behind our previous studies 
%  \cite{Bonelli:2005ti, Bonelli:2005rw, Bonelli:2006ph}. However we have to admit that it was quite 
%   hard to advance beyond the formal construction of the corresponding TFT. 
Lie-algebra description is given by
 \beq
  \frac{dX^\mu}{dt} + \frac{1}{2} \phi^\mu_{~\nu\rho}~ [X^\nu, X^\rho] = 0~,
 \eeq{octoNahm}
   where $\phi$ is the 3-form associated to the $G_2$ structure on $\mathbb{R}^7$. 
%There is nothing known about the solution of this equation. 
The equation above can be derived in a similar way as for equation 
(\ref{standardNahm})
from the octonion self-duality equations for a gauge 
field in eight dimensions which were studied in \cite{Baulieu:1997em,Acharya:1997gp,Blau:1997pp,Bonelli:1999it}. 
Analogously the topological matrix theory can be derived as a dimensional reduction of
the eight dimensional topological gauge theories in \cite{Baulieu:1997em,Acharya:1997gp,Blau:1997pp}.

Another possible construction can be obtained by considering the holomorphic analog of Nahm equation 
using the holomorphic volume form for a Calabi-Yau manifold. This would provide a 2d topological theory.

\subsection{$3$-algebras and M2-M5 system}

For $p=3$ we can consider a topological theory of maps
$X:S^1\times \Sigma_3\to \mathbb{R}^4$ with cross vector product 
induced by the volume form $\epsilon_{\mu\nu\rho\sigma}$.
This localizes on maps
\beq
  \d_0 X^\mu +  \epsilon^\mu_{~\nu\rho\sigma}~  \d_1 X^\nu  \d_2 X^\rho \d_3 X^\sigma = 0~.
\eeq{beh}
The dicretized theory then localizes on the solutions to the Basu-Harvey equation
 \beq
  \frac{dX^\mu}{dt} + \frac{1}{3!} \epsilon^\mu_{~\nu\rho\sigma}~ [X^\nu, X^\rho, X^\sigma] = 0~. 
 \eeq{bah}

A generalization of this set-up is obtained by considering the following M2-M5 system.
Consider M-theory on $\mathbb{R}^{1,2}\times X_8$, with $X_8$ an 8-manifold with 
Spin(7) holonomy. If $X_8$ is compact, this geometry produces ${\cal N}=1$ susy in three dimensions.
Let us take a set of M2-branes at points in $X_8$, therefore elongating along $\mathbb{R}^{1,2}$, 
and an M5 on which these membranes terminates. If $C_4$ is any Cayley 4-cycle in $X_8$ containing the points at which 
the M2-branes are located, then the M5-brane can be taken along $\mathbb{R}^{1,1}\times C_4$, where 
$\mathbb{R}^{1,1}\subset \mathbb{R}^{1,2}$
describes the string along the M-branes intersection.

We can consider the system above from two different points of view, namely that of the M2-branes and that of the 
M5-branes. Our aim is to show the connection between the two\footnote{A relation between the M2 and the M5 worldvolume theories 
has been noticed also in \cite{matsuo,Park:2008qe}.}.

Let us consider the system from the M2-branes point of view. This is described by a 
 generalized Bagger-Lambert-Gustavsson theory.
The BPS configurations are obtained by saturating the supersymmetry via the natural spinor provided by the 
Spin(7) holonomy, which satisfies 
$\Psi_{IJKL}\gamma^{IJKL}\varepsilon=\varepsilon$, where $\Psi_{IJKL}$ is the Cayley four-form.
The BPS condition on the supersymmetry variation with zero gauge field reads
\beq
\delta\psi=\partial_\mu X^I \Gamma^\mu_I\epsilon + \Gamma_{JKL}[X^J,X^K,X^L] \epsilon=0~,
\eeq{ecch}
where here $\mu=0,1,2$ are the world-volume indices and $I,J,\ldots=1,\ldots ,8$ the transverse target space ones.  
The corresponding spinor satisfies both the world-volume chirality constraint $\Gamma^{012}\epsilon=\epsilon$
and the $Spin(7)$ polarization $\Gamma^{12IJLK}\Psi_{IJKL}\epsilon=\epsilon$.
We solve them by splitting $Spin(1,10)\to Spin(1,2)\times Spin(8)$ as
$\epsilon=\eta\otimes\varepsilon$, where 
\footnote{$\Gamma^A$ (A=0,...,10) correspondingly split as $\Gamma^\mu=\gamma^\mu\otimes 1_8$ 
and $\Gamma^I=\gamma^{012}\otimes\gamma^I$} $\eta=\gamma^{012}\eta$.
Plugging this in the BPS condition, and using 
$\Psi_{IJKL}\gamma^{IJKL}\gamma^{MNP}\varepsilon=\Psi^{MNP}_L\gamma^L\varepsilon$
(which follows from the chirality of $\varepsilon$ and the self-duality of the $\Psi$ tensor)
we stay with 
\beq
\partial_\mu X^I \gamma^\mu_I\epsilon + \Psi_{JKLI}[X^J,X^K,X^L] \gamma^{12}\gamma^I \epsilon=0 ~.
\eeq{occh}
If we choose now $X^I$ to depend only on the coordinate $t$ transverse to the M5,
by using the world-volume chirality we get
\beq
\left(\partial_0 X^I + \Psi_{JKLI}[X^J,X^K,X^L]\right)\gamma^0\gamma^I \epsilon=0~,
\eeq{akaksooo"}
which implies the (generalized) Basu-Harvey equation \cite{Berman:2006eu,Krishnan:2008zm}
\be
\partial_0 X^I + \Psi_{JKLI}[X^J,X^K,X^L]=0~.
\label{B}
\ee
We observe that for consistency of the $3$-algebra theory $\Psi_{JKLI}$ should undergo
some constraints coming from the integrability condition of (\ref{B}) which are solved
by flat $X^8$ and constant $\Psi$.
     
Let us now take the point of view of the euclidean M5-brane. 
This is located at some given value of a spatial coordinate of the M2 brane, say $X^1$, and wraps the 
Cayley four cycle.
Its world-volume is then $R^{(1,1)}\times C_4$.
From the internal $Spin(7)$ manifold point of view the M5 brane corresponds to an euclidean 
three-brane wrapping a Cayley submanifold. A topological theory counting such BPS
objects was defined in \cite{Bonelli:2005rw}. This corresponds to a theory of maps
embedding the world-volume of the three-brane $S^1\times\Sigma_3$ into a $Spin(7)$ manifold 
satisfying the condition
\be
\partial_0 X^I+\Psi^{IJKL} \partial_1 X^J \partial_2 X^K \partial_3 X^L = 0
\label{A}
\ee
where $\partial_0$ is the derivative in the 
$S^1$ direction and the remaining are along the $\Sigma_3$ manifold.
By a suitable gauge-fixing of the M5-brane world-volume diffeomorphisms
we can identify this direction with $t$.
The above equations can be regarded as the large N limit of
(generalized) Basu-Harvey equations!
Indeed, following the results of the previous section, by using the Nambu bracket on $\Sigma_3$
%\be
%\{X^J, X^K, X^L \}\equiv * dX^J \wedge dX^K \wedge dX^L 
%\label{nambu}
%\ee
%where $*$ is the Hodge-star operator, 
we can rewrite (\ref{A}) as
\be
\partial_0 X^I+\Psi^{IJKL} \{X^J, X^K, X^L \} = 0
\label{A'}
\ee
%Then the relation between Nambu brackets and three-algebrae 
%in the large N limit
%\be
%[X^J,X^K,X^L] \to  * dX^J \wedge dX^K \wedge dX^L 
%\ee
%immediately 
implying that (\ref{B}) reproduces (\ref{A'}) in the continuum limit.

 \section{Further directions}
 \label{end}

In this note we have discussed that the truncation of Nambu algebras to $p$-algebras
 gives rise to  a novel class of topological (quantum mechanical) theories
which can be regarded as regularizations of topological brane theories.
Moreover, this truncation provides  a link between  $p$-brane instantons and solutions of 
generalized Nahm equations.
The generalized Nahm equation
$$
\dot{X}^\mu + \frac{1}{p!} \phi^\mu_{~~{\nu_1}...{\nu_p}} [X^{\nu_1}, ... , X^{\nu_p} ]=0
$$
can be studied with the natural boundary conditions
$X^\mu(t)\sim\frac{X^\mu_0}{t^{\frac{1}{p-1}}}$ up to less singular terms,
where $X^\mu_0$ is an element in the $p$-algebra satisfying the relation
$$
X^\mu_0 = \frac{1}{p (p-2)!} \phi^\mu_{~~{\nu_1}...{\nu_p}} [X_0^{\nu_1}, ... , X_0^{\nu_p} ]~.
$$
This could be taken as a starting point to define generalized fuzzy geometries related to $p$-brane instantons.

An application of the $p$-algebra TFT that we elaborated could be the definition of  
a regularized version of the moduli space of supersymmetric calibrated cycles in terms of 
algebraic equations. 
Indeed, the limited knowledge of the properties of these spaces is one of the main obstructions 
to fully explore the topological brane theory at quantum level. 
It is crucial in this respect that the regularization is preserving the topological features 
of the theory.

It could be interesting also to investigate whether our $p$-algebra instanton equation provides an
ADHMN-like construction of other BPS configurations in the Bagger-Lambert theories (and its massive deformations)
as for example Chern-Simons vortices (see \cite{Hosomichi:2008qk} too for some related comments).

\noindent{\bf Acknowledgment}:  We are grateful to
M. Cirafici,  A. Kashani-Poor and  C. Maccaferri
  for discussions. 
  The research of G.B. has been supported by MIUR
under the program ``Teoria dei Campi, Superstringhe e Gravit\`a'' and by
the European Commission RTN Program MRTN-CT-2004-005104.
 M.Z. thanks  KITP, Santa Barbara
   where part of this work was carried out. 
 The research of M.Z. was supported by  VR-grant 621-2004-3177,
  in part by the National Science Foundation under Grant No. NSF PHY05-51164 
   and in part  by DARPA and AFOSR through the grant FA9550-07-1-0543.

\appendix 

\Section{Vector cross product structure}
\label{VCP}

 In this Appendix we review the real and complex vector cross product structures on smooth manifolds. 

We start from the real version of vector cross product.  The real vector product 
 is an extension of the standard vector cross product $\times$ of two vectors in $\mathbb{R}^3$ to 
  a smooth manifolds with a metric.
  The generalization of vector cross product to a Riemannian manifold
  leads to the following definition by Brown and Gray \cite{brown}

\begin{definition}
 On a $d$-dimensional Riemannian manifold $M$ with metric $g$ a $p$-fold vector cross product
 is defined by a smooth bundle map
$$  \chi : \wedge^p TM \rightarrow TM $$
satisfying
$$g (\chi(v_1,...,v_p), v_i)=0,\,\,\,\,\,\,1 \leq i \leq p$$
$$ g(\chi(v_1,...,v_p), \chi(v_1,...,v_p)) = \| v_1 \wedge ... \wedge v_p \|^2$$
 where $\|...\|$ is the induced metric on $\wedge^p TM$.
\end{definition}
 Equivalently the last property can be rewritten in the following form
$$ g(\chi(v_1,...,v_p), \chi(v_1,...,v_p)) = \det ( g(v_i, v_j))= \| v_1 \wedge ... \wedge v_p \|^2. $$
 The first condition in the above definition is equivalent to the following tensor $\phi$
$$ \phi (v_1,...,v_p, v_{p+1}) = g (\chi(v_1,...,v_p), v_{p+1})$$
 being a skew symmetric tensor of degree $p+1$, i.e. $\phi \in \Omega^{p+1}(M)$. 
 Thus through the paper we refer to the $(p+1)$-form $\phi$ as $p$-fold vector 
 cross product $(\phi, g)$.  Alternatively a vector cross product form can be defined via a form $\phi \in \Omega^{p+1}(M)$
  satisfying the following property
  $$ \| i_{e_1 \wedge e_2 \wedge ...\wedge e_{p}} \phi \| =1$$
  for any orthonormal set $e_1, e_2, ..., e_p \in T_x M$ and any $x \in M$.

Cross products on real spaces  were classified by Brown and Gray \cite{brown}. 
 The global vector cross products on manifolds were first studied by Gray \cite{gray}.
 They fall into four categories:

(1) $p=d-1$ and $\phi$ is the volume form of the manifold.

(2) $d$ is even and $p=1$. In this case we have a one-fold cross product $J: TM \rightarrow TM$. 
 Such a map satisfies $J^2=-1$ and is an almost complex structure. The associated $2$-form 
 is the K\"ahler form $\omega = gJ$.

(3) The first of two exceptional cases is a $2$-fold cross product ($p=2$) on a $7$-manifold.
 Such a structure  is called a $G_2$-structure and the associated $3$-form is called 
 a $G_2$-form. 

(4) The second exceptional case is $3$-fold cross product ($p=3$) on an
 $8$-manifold. This is called a $Spin(7)$-structure and the associated 
 $4$-form is called $Spin(7)$-form.

  The complex version of vector cross product has been introduced in \cite{complex}.
   Consider a Hermitian manifold $(g, J, M)$ and define the complex vector cross 
    product as a holomorphic $(p+1)$-form satisfying 
    $$ \| i_{e_1 \wedge e_2 \wedge ...\wedge e_{p}} \phi \| =2^{(p+1)/2}$$
for any orthonormal tangent vectors $e_1, e_2,...,e_p \in T^{1,0}_xM$, for any $x\in M$.
One can show from this definition that $\phi$ can be either a holomorphic symplectic form
 or a holomorphic volume form \cite{complex}.  Thus the examples of manifolds equipped with
  the complex vector cross product structure are hyperk\"ahler and Calabi-Yau manifolds.

\Section{Nambu bracket and $p$-algebra}
\label{Nambu}

In this Appendix we review some standard facts about the Nambu brackets and $p$-algebras
 (Nambu-Lie algebras). The Nambu bracket was proposed  in \cite{Nambu:1973qe}. The first 
 systematic study of it and its Lie algebra analog has been done by L.~Takhtajan
 in \cite{Takhtajan:1993vr}.  In the present review we follow closely this original work. 
  For further applications of the classical and quantum Nambu brackets the reader may consult \cite{Curtright:2002fd, Zachos:2003as}. 
 
 The associative algebra $A$ is equipped with a Nambu bracket of order $p$
 $$ \{~,~,~...~,~\}~:~ A \otimes A \otimes ... \otimes A \longrightarrow A   $$
 which satisfies the following properties for all elements of $A$ 
 
\noindent
 $\bullet$ skew-symmetry 
 $$\{f_1, ..., f_p\} = (-1)^{\epsilon(\sigma)} \{ f_{\sigma(1)}, ... , f_{\sigma(p)}\}$$
 $\bullet$ Leibniz rule 
 $$\{f_1 f_2, f_3, ..., f_{p+1}\}
  = f_1 \{  f_2, f_3, ..., f_{p+1}\} + \{f_1, f_3, ... , f_{p+1}\} f_2 $$
 $\bullet$ fundamental identity
 $$\{ \{ f_1, ... , f_{p-1}, f_p\}, f_{p+1}, ... , f_{2p-1}\} + \{f_p,  \{ f_1, ... , f_{p-1}, f_{p+1}\}, f_{p+2}, ... , f_{2p-1}\}  + ... $$
$$  +  \{ f_p, ... , f_{2p-2}, \{ f_1, ... , f_{p-1}, f_{2p-1}\}\}  = \{ f_1, ... , f_{p-1} \{ f_p, ... , f_{2p-1}\} \} $$ 
For the case of $p=2$ this is the well-known notion of Poisson algebra. The associative algebra $A$ 
 with such a bracket can be called a Nambu algebra. 

If $A = C^\infty (\Sigma)$ is an algebra of smooth functions with point-wise 
 multiplication on a smooth manifold $\Sigma$  
 then Nambu bracket bracket of the functions is defined  as above. The canonical example 
  of a Nambu bracket of order $p$ on an oriented $p$-dimensional manifold is given by a Jacobian (volume 
   form)
   $$ \{f_1, ... , f_p\} = \frac{\d(f_1, ... , f_p)}{\d(\sigma_1, ... , \sigma_p)} = \epsilon^{a_1 ... a_p}~\d_{a_1} f_1~ ...~ \d_{a_p} f^p~.$$
    The notion of Nambu bracket of order $p$  is quite rigid since it contains an infinite family 
     of subordinated Nambu brackets of lower order with the matching condition between them.
      Namely for a fixed $H\in A$ we can define a new bracket 
      $$ \{f_1, ..., f_{p-1}\}_H = \{ H, f_1, ... , f_{p-1}\}~,$$
       which turns out to be a Nambu bracket of order $p-1$. 
       
  On the vector space $V$ the linear Nambu brackets 
   are related to  the notion of  (Nambu-Lie) $p$-algebra.
   (Nambu-Lie) p-algebra is a 
  vector space equipped with a linear skew-symmetric bracket
  $$ [~,~,~...~,~]~:~ V \otimes V \otimes ... \otimes V \longrightarrow V~,   $$
   which satisfies the fundamental identity 
   $$[[ v_1, ... , v_{p-1}, v_p], v_{p+1}, ... , v_{2p-1}] + [v_p,  [ v_1, ... , v_{p-1}, v_{p+1}], v_{p+2}, ... , v_{2p-1}]  + ... $$
$$  +  [ v_p, ... , v_{2p-2}, [ v_1, ... , v_{p-1}, v_{2p-1}]]  = [ v_1, ... , v_{p-1} [ v_p, ... , v_{2p-1}]] ~.$$
 In the case of $p=2$ this is just standard Jacobi identity. 
 If we introduce the basis $T^a$ of $V$ then $p$-bracket can be defined through the structure
  constants
  $$ [T^{a_1}, T^{a_2},  ... , T^{a_p}] = f^{a_1 a_2 ... a_p}_{~~~~~~~~b} T^b$$
  and the fundamental identity becomes
  $$  f^{a_1 ... a_{p-1} a_{p}}_{~~~~~~~~~~~c}~ f^{c a_{p+1}  ... a_{2p-1}}_{~~~~~~~~~~~~~b} +
  f^{a_1 ... a_{p-1}a_{p+1}}_{~~~~~~~~~~~~~~c}~ f^{a_p c a_{p+2} ... a_{2p-1}}_{~~~~~~~~~~~~~~~b} +  ... +
  f^{a_1 ... a_{p-1}  a_{2p-1}}_{~~~~~~~~~~~~~~c} ~f^{a_p ... a_{2p-2}c}_{~~~~~~~~~~~b} = $$
  $$= f^{a_p ... a_{2p-1}}_{~~~~~~~~~~c}~ f^{a_1 ... a_{p-1}c}_{~~~~~~~~~~~b}~. $$
 The
 adjoint action of $\wedge^{p-1}V$ on $V$ is defined as follows
  $$ ad_\Lambda v =   f^{a_1 a_2 ... a_p}_{~~~~~~~~~b} ~\Lambda_{a_1 .. a_{p-1}} v_p ~T^b~, $$
   where  $v= v_a T^a$ and $\Lambda = \Lambda_{a_1 .. a_{p-1}}  T^{a_1} \wedge ... \wedge T^{a_{p-1}}
    \in \wedge^{p-1} V$.  The fundamental identity is equivalent to the statement that 
     the adjoint action acts as a derivation on the bracket
     $$ ad_\Lambda \left ( [ v_1, ... , v_p] \right ) = [ (ad_\Lambda v_1), ... , v_p] + ... + [v_1, ... , (ad_\Lambda v_p)]~.$$
  The derivations $ad_\Lambda$ obviously form a Lie algebra
  $$ ad_\Lambda ad_{\tilde{\Lambda}} - ad_{\tilde{\Lambda}} ad_{\Lambda} = ad_{[[\Lambda, \tilde{\Lambda}]]}~,$$
   with the bracket which we denote $[[~,~]]$.  
   
 The $p$-algebra $V$ is equipped with the invariant trace (non-degenerate metric on $V$) 
 $$ h^{ab} = Tr (T^a, T^b)~,$$
 if it is invariant under the adjoint action, i.e. 
 $$ Tr ( v, ad_\Lambda w) = - Tr (ad_\Lambda v , w)~.$$ 
  This implies that $f^{a_1...a_p b} = h^{bc} ~f^{a_1...a_p}_{~~~~~~~c}$  is totally antisymmetric.  
   Assuming the positivity of metric  $h$ leads to severe 
    restrictions on the structure constants of $p$-algebra \cite{papa}. 
  
  The adjoint action of a $p$-algebra with an invariant metric can be described alternatively 
   through the matrix action on  $V$. 
  The element $\Lambda \in \wedge^{p-1} V$ can be 
    mapped to $Mat_{n\times n}$, $n= \dim V$ as follows
    $$  \lambda^c_{~b}= f^{a_1 a_2 ... a_{p-1}c}_{~~~~~~~~~~~~~b} ~\Lambda_{a_1 .. a_{p-1}}~,$$
     such that $\lambda$'s satisfy the following properties 
     $$   f^{a_1...a_p}_{~~~~~~c} ~\lambda^{c}_{~b} =
      f^{ca_2 ... a_p}_{~~~~~~~~b} ~\lambda^{a_1}_{~c} +
      ~...~ + f^{a_1 ... a_{p-1}c}_{~~~~~~~~~~b} ~ \lambda^{a_p}_{~c} ~, $$
     $$  \lambda^{a}_{~c} h^{cb} = - h^{ac} \lambda^{b}_{~a}  ~.$$

%\vspace{-1cm} 

\end{document}